\documentclass[amsmath,amssymb,showpacs,preprint]{revtex4}
\usepackage{epsfig}
\usepackage{graphicx}

\begin{document}

\title{Finslerian perturbation for the $\Lambda$CDM model}

\author{Xin Li}
\email{lixin@ihep.ac.cn}
\author{Sai Wang}
\email{wangsai@ihep.ac.cn}
\author{Zhe Chang}
\email{changz@ihep.ac.cn}
\affiliation{Institute of High Energy Physics,\\
Theoretical Physics Center for Science Facilities,\\
Chinese Academy of Sciences, 100049 Beijing, China}

\begin{abstract}
We present Finslerian perturbation for the $\Lambda$CDM model, which breaks the isotropic symmetry of the universe.  The analysis on the Killing vectors shows that the Randers-Finsler spacetime breaks the isotropic symmetry even if the scalar perturbations of the FRW metric vanish. In Randers-Finsler spacetime, the modified geodesic equation deduces a modified Boltzmann equation. We propose a perturbational version of the gravitational field equation in Randers-Finsler spacetime, where we have omitted the curvature tensor that does not belong to the base space of the tangent bundle. The gravitational field equations for the gravitational wave are also presented. The primordial power spectrum of the gravitational wave is investigated. We show that the primordial power spectrum for super-horizon perturbations is unchanged. For sub-horizon perturbations, however, the power spectrum is modified.
\end{abstract}

\maketitle
\section{Introduction}
The standard cosmological model, i.e., the $\Lambda$CDM model \cite{Sahni,Padmanabhan} has been well established.  It is based on some basic assumptions \cite{Perivolaropoulos}. One of the most important assumptions is that all the components of the universe, such as matter and dark energy, participate gravitational interaction, which follows the laws of general relativity. The other important assumption states that our universe is homogeneous and isotropic on large scales.

During the past decades, Wilkinson Microwave Anisotropy Probe (WMAP) \cite{Komatsu} provided precise cosmological observations for the cosmic microwave background (CMB). Most of them are consistent with the $\Lambda$CDM model. Recently, the data of Planck Collaboration \cite{Planck1} was released. The Planck satellite observed the CMB with much higher precision. Once again, most of them are consistent with the $\Lambda$CDM model except for some anomalous phenomena. An incomplete but succinct list includes: low-$l$ controversy \cite{C2}, the hemispherical asymmetry \cite{hemisperial as}, the parity asymmetry \cite{parity as}, the quadrupole-octopole alignment \cite{quad align}. Besides observations for CMB, other astronomical observations also show some anomalous behaviors that can not be explained by the $\Lambda$CDM model. For example, by analyzing the Union2 SnIa data, Antoniou {\it et al}. \cite{Antoniou} showed that a preferred direction $(l,b)=(309^\circ,18^\circ)$ in galactic coordinate system for the maximum acceleration of the universe. By analyzing the quasar absorption spectra, Webb {\it et al}. \cite{Webb} showed the fine structure constant varies at large scales. These facts hint that our universe has a preferred direction, which means that the universe is anisotropic.

Finslerian gravity is one possible model that provides naturally an anisotropic spacetime. It is a covariant gravitational theory that based on Finslerian spacetime. Finsler geometry \cite{Book by Bao} is a new geometry which involves Riemann geometry as its special case. Prof. S.S. Chern has pointed out that Finsler geometry is just Riemann geometry without the quadratic restriction, in his Notices of AMS, September, 1996. The symmetry of spacetime is described by the so called isometric group. The generators of isometric group are directly connected  with the Killing vectors. It is well known that the isometric group is a Lie group in Riemannian manifold. This fact also holds in Finslerian manifold\cite{Deng}. Generally, Finslerian spacetime admits less Killing vectors than Riemannian spacetime does\cite{Finsler PF}. The numbers of independent Killing vectors of a $n$ dimensional non-Riemannian Finslerian spacetime should no more than $\frac{n(n-1)}{2}+1$ \cite{Wang}. In general, Finsler spacetime breaks symmetry of spacetime, which naturally involves a preferred direction.

Models based on Finsler spacetime have been proposed. An incomplete but succinct list includes: the Finslerian line element \cite{Bogoslovsky} shows symmetry of Very Special Relativity \cite{Coleman,Gibbons}, the spatial variation of fine structure constant \cite{Webb} could be explained by the anisotropic redshift in Finsler spacetime \cite{Fine structure}, the observations of Bullet Cluster is well described by Finslerian gravity model \cite{Finsler Bullet}, Finslerian perturbation for the primordial power spectra \cite{Finsler inflation}.

In this paper, we consider Finslerian perturbation for the $\Lambda$CDM model.
Certainly, our model should be constrained by the precise cosmological observations, such as WMAP \cite{Komatsu} and Planck satellite \cite{Planck1}. Before doing the numerical calculations, we should first establish a theoretical description of the Finslerian perturbation for the $\Lambda$CDM model.

This paper is organized as follows. Sec.II is divided into three subsections. In subsection A, we study the symmetry of Randers-Finsler spacetime. In subsection B, the geodesic equation in Randers-Finsler spacetime is presented. The Boltzmann equation is given. In subsection C, we discuss the gravitational field equation in Randers-Finsler spacetime, and propose a perturbational version of it. In Sec.III, we present the gravitational field equation of the gravitational wave. Then, we use it to study the primordial power spectrum of the gravitational wave in Randers-Finsler spacetime. In Sec IV, we give conclusions and remarks.

In this paper, notations and convention follow the book written by Dodelson \cite{Dodelson}.

\section{Finslerian perturbation for the $\Lambda$CDM model}
\subsection{Symmetry of Randers-Finsler spacetime}
Instead of defining an inner product structure over the tangent bundle in Riemann geometry, Finsler geometry is based on
the so called Finsler structure $F$ with the property
$F(x,\lambda y)=\lambda F(x,y)$ for all $\lambda>0$, where $x\in M$ represents position
and $y\equiv\frac{dx}{d\tau}$ represents velocity. The Finslerian metric is given as\cite{Book
by Bao}
\begin{equation}
g_{\mu\nu}\equiv\frac{\partial}{\partial
y^\mu}\frac{\partial}{\partial y^\nu}\left(\frac{1}{2}F^2\right).
\end{equation}
Two types of Finsler space should be noticed. One is the Riemann space. A Finslerian metric is said to be Riemannian, if $F^2$ is quadratic in $y$. Another is Randers spacetime \cite{Randers}. It is given as
\begin{equation}\label{Randers form}
F(x,y)\equiv \alpha(x,y)+\beta(x,y),
\end{equation} where
\begin{eqnarray}
\alpha (x,y)&\equiv&\sqrt{\tilde{a}_{\mu\nu}(x)y^\mu y^\nu},\\
\beta(x,y)&\equiv& \tilde{b}_\mu(x)y^\mu,
\end{eqnarray} and $\tilde{a}_{ij}$ is Riemannian metric. Throughout this paper, the indices are lowered and raised by $g_{\mu\nu}$ and its inverse matrix $g^{\mu\nu}$. And the objects that decorate with a tilde are lowered and raised by $\tilde{a}_{\mu\nu}$ and its inverse matrix $\tilde{a}^{\mu\nu}$.

To investigate the Killing vectors, we should construct the isometric
transformations of Finsler structure. It is convenient to
discuss the isometric transformations under an infinitesimal
coordinate transformation for $x$
\begin{equation}
\label{coordinate tran}
\bar{x}^\mu=x^\mu+\epsilon \tilde{V}^\mu,
\end{equation}
together with a corresponding transformation for $y$
\begin{equation}
\label{coordinate tran1}
\bar{y}^\mu=y^\mu+\epsilon\frac{\partial \tilde{V}^\mu}{\partial x^\nu}y^\nu,
\end{equation}
where $|\epsilon|\ll1$.
Under the coordinate transformation (\ref{coordinate tran}) and (\ref{coordinate tran1}), to first order in $|\epsilon|$, we obtain the expansion of the Finsler structure,
\begin{equation}
\label{coordinate tran F}
\bar{F}(\bar{x},\bar{y})=\bar{F}(x,y)+\epsilon \tilde{V}^\mu\frac{\partial F}{\partial x^\mu}+\epsilon y^\nu\frac{\partial \tilde{V}^\mu}{\partial x^\nu}\frac{\partial F}{\partial y^\mu},
\end{equation}
where $\bar{F}(\bar{x},\bar{y})$ should equal to $F(x,y)$.
Under the transformation (\ref{coordinate tran}) and (\ref{coordinate tran1}), a Finsler structure is called isometry if and only if
\begin{equation}
F(x,y)=\bar{F}(x,y).
\end{equation}
Deducing from the (\ref{coordinate tran F}), we obtain the Killing equation $K_V(F)$ in Finsler space
\begin{equation}
\label{killing F}
K_V(F)\equiv \tilde{V}^\mu\frac{\partial F}{\partial x^\mu}+y^\nu\frac{\partial \tilde{V}^\mu}{\partial x^\nu}\frac{\partial F}{\partial y^\mu}=0.
\end{equation}

Plugging the length element of Randers spacetime (\ref{Randers form}) into the Killing equation (\ref{killing F1}), and noticing that the rational and irrational parts of Killing equation are independent, we obtain that
\begin{eqnarray}
\label{killing F1}
\tilde{V}_{\mu|\nu}+\tilde{V}_{\nu|\mu}&=&0,\\
\label{killing F2}
\tilde{V}^\mu\frac{\partial \tilde{b}_\nu}{\partial x^\mu}+\tilde{b}_\mu\frac{\partial \tilde{V}^\mu}{\partial x^\nu}&=&0,
\end{eqnarray}
where $``|"$ denotes the covariant derivative with respect to the Riemannian metric $\alpha$. It is obvious that Killing equation (\ref{killing F1}) is the same as the Riemannian one. But, the other Killing equation (\ref{killing F2}) obtained constrains the first one (\ref{killing F1}). Thus, the number of independent killing vectors in Randers-Finsler spacetime (\ref{Randers form}) is less than the one in Riemannian spacetime $\alpha$ \cite{Finsler PF}.

In this paper, we study the Finslerian perturbation for the $\Lambda$CDM model. So, we take the Riemannian metric $\tilde{a}_{\mu\nu}$ to be the Friedmann-Robertson-Walker (FRW) metric \cite{Weinberg} with scalar perturbations \cite{Dodelson}. The non vanishing components of $\tilde{a}_{\mu\nu}$ are given as
\begin{eqnarray}\label{CMB a00}
\tilde{a}_{00}&=&1+2\psi(\vec{x},t),\\
\label{CMB aij}
\tilde{a}_{ij}&=&-R^2(t)(1+2\phi(\vec{x},t)),
\end{eqnarray}
where $R(t)$ denotes cosmic scale factor, and $\psi$ and $\phi$ are scalar perturbations for the FRW metric. We set the Finslerian perturbation parameter $\tilde{b}_{\mu}$ to be the form 
\begin{equation}\label{one form}
\tilde{b}_\mu=\{b(\vec{x},t),0,0,0\}.
\end{equation}
The scalar perturbations $\psi$ and $\phi$ and the Finslerian perturbation $\tilde{b}$ are first order quantities. In the absence of these perturbations, the Randers metric returns to the FRW metric. It is well known that the FRW metric is spatially homogenous and isotropic, and the scalar perturbations $\psi$ and $\phi$ break isotropy. While the scalar perturbations $\psi$ and $\phi$ vanish, the Finslerian perturbation (\ref{one form}) still breaks the isotropic symmetry of FRW metric. Plugging the FRW metric into the Killing equations (\ref{killing F1}) and (\ref{killing F2}), we obtain that
\begin{eqnarray}\label{solu1}
\tilde{V}^{i}&=&Q^i_{~j} x^j+C^i,\\\label{solu2}
0&=&(Q^i_{~k} x^k+C^i)\frac{\partial \tilde{b}}{\partial x^i},
\end{eqnarray}
where $Q_{ij}=\delta_{ik}Q^k_{~j}$ is an arbitrary constant skew-symmetric matrix and $C_i=\delta_{ik}C^k$ is an arbitrary constant vector. The solution (\ref{solu1}) just represents spatially homogenous and isotropic for FRW metric. Another solution (\ref{solu2}) can not be satisfied unless $\tilde{b}$ is not function of $x^i$. Thus, even if the scalar perturbations $\psi$ and $\phi$ vanish, the Finsler perturbation breaks the isotropic symmetry of FRW metric. Also, one should notice that it breaks the parity symmetry if $b(\vec{x},t)\neq b(-\vec{x},t)$.

\subsection{Geodesic equations in Rander-Finsler spacetime}
In the thermal history of the universe \cite{Mazumdar}, elementary particles and other components of the universe, tightly couple to each other. This could be described by the Boltzmann equation
\begin{equation}\label{Boltzmann eq}
\frac{df}{dt}=C[f],
\end{equation}
where $f$ denotes the distribution function for the components of the universe and $C[f]$ denotes all possible collision terms. Up to first order in perturbation, the left-hand side of Boltzmann equation (\ref{Boltzmann eq}) can be written as
\begin{equation}\label{Boltzmann eq1}
\frac{df}{dt}=\frac{\partial f}{\partial t}+\frac{\partial f}{\partial x^i}\frac{\partial x^i}{\partial t}+\frac{\partial f}{\partial E}\frac{\partial E}{\partial t},
\end{equation}
where $E\equiv(1+\psi)\frac{dx^0}{d\tau}$. The term $\frac{df}{dt}$ directly depends on the velocities of particles $\frac{\partial x^i}{\partial t}$ and $\frac{\partial E}{\partial t}$ that depends on the geodesic equation. Since $\frac{\partial f}{\partial x^i}$ is a first order quantity, we can neglect the first order quantity in $\frac{\partial x^i}{\partial t}$. Then, it is the same with the one in the FRW cosmology
\begin{equation}\label{dx dt}
\frac{\partial x^i}{\partial t}=\frac{n^ip}{RE},
\end{equation}
where $p=\sqrt{E^2-m^2}$ ($m$ is the mass of particles) and $n^i$ is a unit direction vector that satisfies $\delta_{ij}n^in^j=1$.

Now, we need to derive the term $\frac{\partial E}{\partial t}$ by using the Finslerian geodesic equation.
The geodesic equation for Finsler manifold is given as\cite{Book by Bao}
\begin{equation}
\label{geodesic}
\frac{d^2x^\mu}{d\tau^2}+G^\mu=\frac{d\log F}{d\tau}\frac{dx^\mu}{d\tau},
\end{equation}
where
\begin{equation}
\label{geodesic spray}
G^\mu=\frac{1}{2}g^{\mu\nu}\left(\frac{\partial^2 F^2}{\partial x^\lambda \partial y^\nu}y^\lambda-\frac{\partial F^2}{\partial x^\nu}\right)
\end{equation} is called geodesic spray coefficient. If the Finslerian geodesic preserves the Finslerian length, which means that $F$ is a constant along the geodesic, the geodesic equation (\ref{geodesic}) simplifies as
\begin{equation}
\label{geodesic1}
\frac{d^2x^\mu}{d\tau^2}+G^\mu=0.
\end{equation}
Obviously, if $F$ is Riemannian metric, then
\begin{equation}
G^\mu=\tilde{\gamma}^\mu_{\nu\lambda}y^\nu y^\lambda,
\end{equation}
where $\tilde{\gamma}^\mu_{\nu\lambda}$ is the Riemannian Christoffel symbol.

Plugging the Randers metric (\ref{Randers form}) into the geodesic equation (\ref{geodesic1}), we find that \cite{Book by Bao}
\begin{equation}
\frac{d^2x^\mu}{d\tau^2}+\left(\tilde{\gamma}^\mu_{\nu\lambda}+\frac{\tilde{b}^\mu}{F}\tilde{b}_{\nu|\lambda}\right)\frac{dx^\nu}{d\tau} \frac{dx^\lambda}{d\tau}+\left(\tilde{a}^{\mu\nu}-\frac{\tilde{b}^\nu}{F}\frac{dx^\mu}{d\tau}\right)(\tilde{b}_{\nu|\lambda}-\tilde{b}_{\lambda|\nu})\alpha \frac{dx^\lambda}{d\tau}=0.
\end{equation}
The collision terms $C[f]$ in the Boltzmann equation involves scattering processes such as Compton scattering. The calculations of these scattering processes require the dispersion relations of the elementary particles. In Finsler spacetime, the dispersion relations are modified \cite{Finsler dispersion,Finsler PF} in high energy scale. However, the typical energy scale of the thermal period of the universe is Mev. Thus, the Finslerian modification of dispersion relations should vanish. Thus, there is no Finslerian perturbation for the collision terms $C[f]$. This result implies that $F$ is not constant along the geodesic, but the Riemannnian length $\alpha$ is. By making use of this result, we obtain from the equation (\ref{geodesic}) that the geodesic equation which preserves Riemannian length element $\alpha$ is given as
\begin{equation}\label{geodesic randers}
\frac{d^2x^\mu}{d\tau^2}+\tilde{\gamma}^\mu_{\nu\lambda}\frac{dx^\nu}{d\tau} \frac{dx^\lambda}{d\tau}+\tilde{a}^{\mu\nu}(\tilde{b}_{\nu|\lambda}-\tilde{b}_{\lambda|\nu})\alpha \frac{dx^\lambda}{d\tau}=0.
\end{equation}
Plugging $E\equiv(1+\psi)\frac{dx^0}{d\tau}$ into the geodesic equation, we obtain
\begin{equation}\label{dE dt}
\frac{dE}{dt}=-\left(\frac{p^2}{E}\left(H+\frac{\partial \phi}{\partial t}\right)+\frac{pn^i}{R}\left(\frac{\partial \psi}{\partial x^i}+\frac{\alpha}{E}\frac{\partial b}{\partial x^i}\right)\right),
\end{equation}
where $H\equiv\frac{\dot{R}}{R}$.
Here, the dot denotes derivative with respect to time. Then, by making use of equations (\ref{dx dt},\ref{dE dt}), we obtain the Boltzmann equations in Randers-Finsler spacetime
\begin{eqnarray}\label{Boltzmann massless}
\frac{\partial f}{\partial t}+\frac{n^i}{R}\frac{\partial f}{\partial x^i}-p\frac{\partial f}{\partial E}\left(H+\frac{\partial \phi}{\partial t}+\frac{n^i}{R}\frac{\partial \psi}{\partial x^i}\right)&=&C[f],\\ \label{Boltzmann massive}
\frac{\partial f}{\partial t}+\frac{n^ip}{RE}\frac{\partial f}{\partial x^i}-\frac{\partial f}{\partial E}\left[\frac{p^2}{E}\left(H+\frac{\partial \phi}{\partial t}\right)+\frac{pn^i}{R}\frac{\partial }{\partial x^i}\left(\psi+b\right)\right]&=&C[f],
\end{eqnarray}
for massless particles and massive particles, respectively. Here, we have used the fact that the Riemannian length $\alpha=0$ for massless particle to deduce the equation (\ref{Boltzmann massless}). And, we have used the fact that the velocity of non-relativistic particles can be regarded as a first order quantity to deduce the equation (\ref{Boltzmann massive}).

It is obvious from (\ref{Boltzmann massless}) that the Boltzmann equation for massless particles in Randers-Finsler spacetime is the same with the one \cite{Dodelson} in the FRW metric with scalar perturbations $\psi$ and $\phi$. One can simply get the Boltzmann equation for massive particles in Randers-Finsler spacetime by transforming $\psi\longrightarrow\psi+b$ from the one in the FRW metric with scalar perturbations.

\subsection{Gravitational field equations in Rander-Finsler spacetime}
In general relativity, the Einstein gravitational field equations decide how the perturbations to the distributions of the comic components affect the scalar perturbations $\psi$ and $\phi$ in the $\Lambda$CDM model. Analogously, we need gravitational field equations in Finsler spacetime to decide the effect of Finslerian perturbation $b$. However, up to now, debate still exists for the gravitational field equation in Finsler spacetime. There are many types of gravitational field equation in Finsler spacetime \cite{Miron,Rutz,Vacaru,Pfeifer}. And these equations do not equal to each other. The Finslerian length element $F$ is constructed on a tangent bundle \cite{Book by Bao}. So, the gravitational field equation should be constructed on the tangent bundle in principle.
However, the corresponded energy-momentum tensor, which should be constructed on the tangent bundle, is rather obscure in physics. In this paper, we just consider a Finslerian perturbation for the $\Lambda$CDM model. So, it is valid on a certain extent to neglect  effects that induced by the tangent bundle.

In Finsler geometry, the curvature $R^{~\mu}_{\lambda~\nu\rho}$ depends on the connection that one chooses. There are many different kinds of connection, such as Berwald connection, Chern connection and Cartan connection \cite{Book by Bao}. If one use the contraction of curvature $R^{~\mu}_{\lambda~\nu\rho}$ to construct a gravitational field equation, it depends on the connection. This is unacceptable in physics.

In Finsler geometry, there is geometrical invariant quantity, i.e., Ricci scalar. It is of the form \cite{Book by Bao}
\begin{equation}\label{predecessor flag curvature}
Ric\equiv R^\mu_{~\mu}=\frac{1}{F^2}\left(\frac{\partial G^\mu}{\partial x^\mu}-y^\lambda\frac{\partial^2 G^\mu}{\partial x^\lambda\partial y^\mu}+2G^\lambda\frac{\partial^2 G^\mu}{\partial y^\lambda\partial y^\mu}-\frac{\partial G^\mu}{\partial y^\lambda}\frac{\partial G^\lambda}{\partial y^\mu}\right),
\end{equation}
where $R^\mu_{~\nu}=R^{~\mu}_{\lambda~\nu\rho}y^\lambda y^\rho/F^2$. Though $R^{~\mu}_{\lambda~\nu\rho}$ depends on connections, $R^\mu_{~\nu}$ does not \cite{Book by Bao}. The Ricci scalar only depends on the Finsler structure $F$ and is insensitive to connections. In Ref. \cite{Finsler Bullet,Finsler GW,Finsler MOND}, we have presented that the vacuum field equation in Finsler gravity is of the form $Ric=0$.
The physical meaning of Ricci scalar is very clear. The Ricci scalar plays an important role in the geodesic deviation equation  \cite{Finsler Bullet,Finsler GW,Finsler MOND,Book by Bao} that is given as
\begin{equation}
-\frac{D^2\xi^\mu}{D\tau^2}=\xi^\nu R^\mu_{~\nu}F^2.
\end{equation}
The vanishing Ricci scalar implies geodesic rays do not bunch together or disperse. It means that it is vacuum outside the gravitational source. The gravitational vacuum field equation $Ric=0$ is universal in any types of theories of Finsler gravity.

In order to construct a gravitational field equation in the presence of energy-momentum source, we need to introduce Ricci tensor in Finsler spacetime. The notion of Ricci tensor in Finsler geometry was first introduced by Akbar-Zadeh\cite{Akbar}
\begin{equation}\label{Ricci tensor}
Ric_{\mu\nu}=\frac{\partial^2\left(\frac{1}{2}F^2 Ric\right)}{\partial y^\mu\partial y^\nu}.
\end{equation}
And the scalar curvature in Finsler geometry is given as $S=g^{\mu\nu}Ric_{\mu\nu}$. Here, we define the modified Einstein tensor in Finsler spacetime
\begin{equation}\label{Einstein tensor}
G_{\mu\nu}\equiv Ric_{\mu\nu}-\frac{1}{2}g_{\mu\nu}S.
\end{equation}
Now, we propose that the gravitational field equation in Finsler spacetime is of the form
\begin{equation}\label{Gravitation eq}
\left(G_{\mu\nu}-8\pi G T_{\mu\nu}\right)_{|M}=0,
\end{equation}
where $T_{\mu\nu}$ is the energy-momentum tensor. Here the `$|M$' means that the gravitational field equation (\ref{Gravitation eq}) restricted to the base space $M$ (four dimensional spacetime) of Finslerian length element $F$, all fiber coordinate $y$ (or velocities) are set to be the velocities of the cosmic components, which are just the same as the fluid velocities of the energy-momentum tensor $T_{\mu\nu}$.

It is obvious that the gravitational field equation (\ref{Gravitation eq}) is insensitive to connections. The equation (\ref{Gravitation eq}) is valid for two reasons. The first one states that the Finslerian parameter $b$ is just a perturbation which means that the Finsler spacetime is very close to the FRW metric. Therefore, the gravitational field equation should possess the same form as it in general relativity. The other states that the gravitational equation (\ref{Gravitation eq}) could be derived by the formulism of Pfeifer {\it et al}. \cite{Pfeifer}. Pfeifer {\it et al}. \cite{Pfeifer} have constructed gravitational dynamics for Finsler spacetimes in terms of an action integral on the unit tangent bundle. Their results show that the gravitational field equation in Finsler spacetime is given as
\begin{equation}\label{Gravitation eq Pfeifer}
S-6Ric+2g^{\mu\nu}\big(\nabla_\mu S_\nu+S_\mu S_\nu+\partial_{y^\mu}\nabla S_\nu\big)=-4\pi G T.
\end{equation}
The $S_\mu$-terms can be written as $S_\mu=y^{\nu}P_{\nu~\lambda \mu}^{~\lambda}/F$, where $P_{\nu~\lambda \mu}^{~\lambda}$ are the coefficients of the cross basis $dx \wedge \frac{\delta y}{F}$ \citep{Book by Bao}. Accordingly, the energy-momentum tensor can also be divided by the basis of $dx\wedge dx$ and $dx \wedge \frac{\delta y}{F}$. Thus, the $S_\mu$-terms contribute to the energy-momentum tensor that belong to the basis $dx \wedge \frac{\delta y}{F}$. As we mentioned at the beginning of this subsection, they can be dropped. Then, it is obvious from the equation (\ref{Gravitation eq Pfeifer}) that $Ric=0$ is one of its solutions for vacuum. The term $T$ in the equation (\ref{Gravitation eq Pfeifer}) could be written as \cite{Pfeifer}
\begin{equation}
T=-2\bar{T}+12\frac{\bar{T}_{\mu\nu}y^\mu y^\nu}{F^2}+f\left(A,\partial_{y^\nu}\frac{\partial \mathcal{L}}{\partial g_{\mu\nu}}\right),
\end{equation}
where $\mathcal{L}$ is Lagrange density of matter action, $\bar{T}_{\mu\nu}=g_{\mu\nu}\mathcal{L}+2\frac{\partial \mathcal{L}}{\partial g^{\mu\nu}}$ possess the same form with its Riemannian one and $\bar{T}=g^{\mu\nu}\bar{T}_{\mu\nu}$. Here, $f$ is a function of Cartan connection \cite{Book by Bao} $A_{\mu\nu\lambda}=F\partial_{y^{\mu}}\partial_{y^{\nu}}\partial_{y^{\lambda}}/4$ and $\partial_{y^\nu}\frac{\partial \mathcal{L}}{\partial g_{\mu\nu}}$. It should be noticed that the derivative of $A$ with respect to $y$ belongs to the coefficients with basis $\frac{\delta y}{F}$. And $\partial_{y}\partial_{y}\mathcal{L}\propto\frac{\partial \mathcal{L}}{\partial g}\partial_{y}A+A\partial_{y}\frac{\partial \mathcal{L}}{\partial g}$ is a second order term. Therefore, calculating the second derivative with respect to $y$ for the equation (\ref{Gravitation eq Pfeifer}), the term that is proportional to $f$ could be neglected. Then one can find that $G_{\mu\nu}=8\pi GT_{\mu\nu}$ is a solution of the equation (\ref{Gravitation eq Pfeifer}). It means that the gravitational field equation (\ref{Gravitation eq}) we proposed is valid.

Plugging the Randers-Finsler length element (\ref{Randers form}) into the formula (\ref{geodesic spray}), then plugging the results into the formula (\ref{predecessor flag curvature}), up to first order in $\psi$ and $\phi$ and $b$, we obtain the Ricci scalar of Randers-Finsler spacetime (\ref{Randers form})
\begin{eqnarray}
F^2Ric=&&Ric_{\mu\nu}^{(R)}y^\mu y^\nu-b_{,i,j}y^0\left(\frac{3y^i y^j}{2\alpha}+\tilde{a}^{ij}\alpha\right)+H \frac{y^ib_{,i}}{\alpha}(4{y^0}^2-2y^j y_j)\nonumber\\
&+&by^jy_j\frac{y^0}{2\alpha}\left(9H^2-3\frac{\ddot{R}}{R}\right)+b_{,0,i}y^i\left(\alpha-\frac{3{y^0}^2}{\alpha}\right)-b_{,0,0}\frac{3{y^0}^3}{2\alpha}-b_{,0}Hy^jy_j\frac{9y^0}{2\alpha},
\end{eqnarray}
where $Ric_{\mu\nu}^{(R)}$ denotes the Ricci tensor of Riemannian metric $\tilde{a}_{\mu\nu}$, $y_j\equiv\tilde{a}_{ij}y^j$, $H\equiv\frac{\dot{R}}{R}$, and the commas and dots denote derivative with respect to $x^\mu$ and $t$, respectively. By making use of the formula (\ref{Ricci tensor}), and restricting the Ricci tensor to the base space $M$ with Finslerian length element $F$, and noticing that the fluid velocities of cosmic components are first order quantities which means the formulae in proportion to $y^i$ in Ricci tensor can be neglected, the Ricci tensors are given as
\begin{eqnarray}\label{Ric 00}
(Ric_{00})_{|M}&=&Ric_{00}^{(R)}-\tilde{a}^{ij}b_{,i,j}-\frac{3}{2}b_{,0,0},\\
\label{Ric oj}
(Ric_{0j})_{|M}&=&Ric_{0j}^{(R)}+2Hb_{,j}-b_{,0,j},\\
\label{Ric ij}
(Ric_{ij})_{|M}&=&Ric_{ij}^{(R)}-\frac{3}{2}b_{,i,j}+\tilde{a}_{ij}\left(\frac{b}{2}\left(9H^2-\frac{3\ddot{R}}{R}\right)-\frac{9Hb_{,0}}{2}+\frac{3b_{,0,0}}{4}-\frac{b_{,m,n}\tilde{a}^{mn}}{2}\right)
\end{eqnarray}
(The following derivations are all calculated on the base space $M$, so we omit the symbol $|M$.)
Then, plugging these equations (\ref{Ric 00},\ref{Ric oj},\ref{Ric ij}) into formula (\ref{Einstein tensor}), we obtain the first order part of $G_{\mu\nu}$, they are given as
\begin{eqnarray}\label{delta G00}
\delta G^0_0&=&6H\phi_{,0}-6\psi H^2-\frac{2\phi_{,k,k}}{R^2}+\tilde{a}^{ij}b_{,i,j}-\frac{15}{8}b_{,0,0}+\frac{27}{4}Hb_{,0}+b\left(\frac{15\ddot{R}}{4R}-\frac{39}{4}H^2\right),\\
\label{delta G0j}
\delta G^0_j&=&-2\phi_{,0,j}+2H\psi_{,j}-b_{,0,j}+2Hb_{,j},\\
\label{delta Gij}
\delta G^i_j&=&\delta {G^i_j}^{(R)}+\delta^i_j\left(-\frac{3b_{,k,k}}{2R^2}-b\left(+\frac{3H^2}{4}+\frac{11\ddot{R}}{4R}\right)+\frac{9H}{4}b_{,0}+\frac{3}{8}b_{,0,0}\right)+\frac{3b_{,i,j}}{2R^2},
\end{eqnarray}
where ${G^i_j}^{(R)}$ denotes the spatial components of Einstein tensor of the Riemannian metric $\tilde{a}_{\mu\nu}$. Since, we have already restricted the gravitational field equation to the base space $M$, the corresponded energy-momentum tensor is just the familiar one to us. It is given as
\begin{equation}\label{energy-momentum}
T^\mu_\nu=\int\frac{d^3p}{(2\pi)^3}\frac{P^\mu P^\nu}{P^0}f,
\end{equation}
where $f$ denotes the distribution function for the cosmic components and $P^\mu\equiv\frac{dx^\mu}{d\tau}$. Thus, the first order part of $T^\mu_\nu$ is the same with that in the $\Lambda$CDM model \cite{Dodelson}.

Now, combining the modified Einstein tensors (\ref{delta G00},\ref{delta G0j},\ref{delta Gij}) and the energy-momentum tensor (\ref{energy-momentum}), we find from the gravitational equation (\ref{Gravitation eq}) that its first order parts in Fourier space (for example, $\phi_{,i}=ik_i\phi$) are given as
\begin{eqnarray}
4\pi G(\rho_m\delta+4\rho_r\Theta_{r0})&=&\frac{k^2\phi}{R^2}+3H(\phi_{,0}-H\psi)\nonumber\\
\label{Gravitation eq G00}
&&+\frac{k^2b}{2R^2}-\frac{15}{8}b_{,0,0}+\frac{27}{4}Hb_{,0}+b\left(\frac{15\ddot{R}}{4R}-\frac{39}{4}H^2\right),\\
\label{Gravitation eq G0j}
\frac{4\pi G R}{i k}(\rho_m v_m-4i\rho_r\Theta_{r1})&=&\phi_{,0}-H\psi+\frac{b_{,0}}{2}-H b,\\
\label{Gravitation eq Gij}
-32\pi G R^2\rho_r\Theta_{r2}&=&k^2\left(\phi+\psi+\frac{3b}{2}\right),
\end{eqnarray}
where $k$ denotes the magnitude of wave vector $k=\sqrt{k_1k_i}$ ($k_i=k^i$), $\rho_m$ and $\rho_r$ denote the energy density of massive particles and massless particles, respectively, $\delta$ and $v_m$ denote fractional overdensity and velocity flows of massive particles, respectively, and $\Theta_{r0},\Theta_{r1},\Theta_{r2}$ denote monopole, dipole and quadrupole of the massless particles distribution, respectively. Here, we just used the non-symmetrical parts of $G^i_j$ to derive the equation (\ref{Gravitation eq Gij}).

\section{Primordial power spectrum of gravitational wave with Finslerian perturbation}
Inflation \cite{Linde} is essential for the $\Lambda$CDM model, which solve the ``horizon problem" and ``flatness problem".
In the former section, we have presented the Finslerian perturbation for the $\Lambda$CDM model in the thermal history of the universe. The Finslerian perturbation for inflation in Randers spacetime have been investigated \cite{Finsler inflation}. During the period of inflation, the typical energy scale is supposed to be very high. Then, the dispersion relation should be modified in Finsler spacetime. It is anticipated that the Finslerian effect should appear in the perturbation of energy-momentum tensor. Indeed, ref. \cite{Finsler inflation} show that the wavenumber $k$ is modified such that the initial power spectrum is of the form $\mathcal{P}(\vec{k})=\mathcal{P}(k)(1+g(\vec{k}\cdot \vec{n}))$ where $n$ denotes a preferred direction of the universe.

In this section, we will give primordial power spectrum of gravitational wave in the given Randers-Finsler spacetime. The non vanishing components of metric are given as
\begin{eqnarray}\label{a00}
\tilde{a}_{00}&=&1,\\
\label{aij}
\tilde{a}_{ij}&=&-R^2(t)(\delta_{ij}+\mathcal{H}_{ij}),\\
\label{b}
\tilde{b}_{\mu}&=&\{b(\vec{x},t),0,0,0\},
\end{eqnarray}
where $\mathcal{H}_{ij}$ contains the perturbation given by the gravitational wave, and is equal to
\begin{eqnarray}\label{gravitational wave}
\mathcal{H}_{ij}=\left(
  \begin{array}{ccc}
    h_+ & h_\times & 0 \\
    h_\times & -h_+ & 0 \\
    0 & 0 & 0 \\
  \end{array}
\right).
\end{eqnarray}
Here, $h_+,h_\times$ and $b$ are assumed small. The formula (\ref{gravitational wave}) means that we have chosen the perturbation to be in $x-y$ plane such that the gravitational wave propagate along $z$-axis. Following the similar process of subsection C in section II, we obtain the first order parts of modified Einstein tensor in the Randers-Finsler spacetime (\ref{a00},\ref{aij},\ref{b}), which are given as
\begin{eqnarray}\label{eq h+}
\delta G^2_2-\delta G^1_1&=&h_{+,0,0}+3H h_{+,0}+\frac{k^2_3}{R^2}h_++\frac{3b}{2R^2}(k_1^2-k_2^2),\\
\label{eq h*}
-\delta G^1_2=-\delta G^2_1&=&\frac{h_{\times.0.0}}{2}+\frac{3}{2}H h_{\times,0}+\frac{k^2_3}{2R^2}h_{\times}+\frac{3b}{2R^2}k_1 k_2,\\
\label{eq G00}
-\delta G^0_0&=&b_{,0,0}-\frac{18}{5}H b_{,0}+\left(\frac{26}{5}H^2-\frac{2\ddot{R}}{R}-\frac{8k^2}{15R^2}\right)b,\\
\label{eq Gij}
-\delta G^i_j&=&\frac{3b}{2R^2}k_i k_j, ~~~(i,j\neq\{1,1\},\{1,2\},\{2,1\},\{2,2\}).
\end{eqnarray}
The gravitational field equation (\ref{Gravitation eq}) requires that the equation (\ref{eq h+}) is proportional to $\delta T^2_2-\delta T^1_1$, and the equation (\ref{eq h*}) is proportional to $-\delta T^1_2$. Only the photon distribution contributes to the anisotropic stress. If its perturbation $\Theta$ depends only on the cosine of the angle between $z$-axis and the direction of photon's momentum, $\delta T^2_2-\delta T^1_1$ equals to $0$. And if $\Theta$ also depends on another angle in spherical coordinate, it will involve a negligible effect to the gravitational wave \cite{Dodelson}. Due to the integration $\int d\Omega n^1 n^2\propto\delta_{12}$ where $\Omega$ denote solid angle and $n^i$ is a unit direction vector, the perturbation to $T^1_2$ is in proportion to $\delta_{12}$, which equals to $0$. Therefore, the gravitational wave in Randers-Finsler spacetime obeys
\begin{eqnarray}\label{eq h+1}
h_{+,0,0}+3H h_{+,0}+\frac{k^2_3}{R^2}h_+ +\frac{3b}{2R^2}(k_1^2-k_2^2)=0,\\
\label{eq h*1}
h_{\times,0,0}+3H h_{\times,0}+\frac{k^2_3}{R^2}h_\times+\frac{3b}{R^2}k_1 k_2=0.
\end{eqnarray}
As was mentioned above, the gravitational wave propagates along the $z$-axis. If the wave vectors of $b$ parallel the direction of propagation of the gravitational wave, it means that $k_1=k_2=0$. Then, the equations (\ref{eq h+1},\ref{eq h*1}) reduce to the one in the $\Lambda$CDM model. This implies that there is no Finslerian effect for the gravitational wave. In order to investigate the Finslerian effect for the gravitational wave, the wave vectors of $b$ must be set perpendicular to the $z$-axis. For simplicity, we choose the wave vector of $b$ to parallel $x$-axis and set $b$ to depend only on the spatial coordinate $\vec{x}$. Then, the equations (\ref{eq h+1},\ref{eq h*1}) reduce to
\begin{eqnarray}\label{eq h+2}
h_{+,0,0}+3H h_{+,0}+\frac{k^2_3}{R^2}h_+ +\frac{3b}{2R^2}k_1^2=0,\\
\label{eq h*2}
h_{\times,0,0}+3H h_{\times,0}+\frac{k^2_3}{R^2}h_\times=0.
\end{eqnarray}
And we also obtain other gravitational field equations from the equations (\ref{eq G00},\ref{eq Gij})that
\begin{eqnarray}\label{GW 00}
\left(\frac{2\ddot{R}}{R}+\frac{8k^2_1}{15R^2}-\frac{26}{5}H^2\right)b&=&8\pi G\delta T^0_0,\\
\label{GW 33}
-\frac{3b}{2R^2}k_1^2&=&8\pi G\delta T^3_3.
\end{eqnarray}
The equations (\ref{GW 00}) and (\ref{GW 33}) just give a Finslerian perturbation to the energy density and stress in the $z$-direction. And the metric perturbation $h_\times$ obeys the same equation with the one in the $\Lambda$CDM model. But the equation for $h_+$ is modified.

Now, we can discuss the role of Finslerian perturbation in the primordial power spectrum of the gravitational wave. By changing the variable of the equations (\ref{eq h+2},\ref{eq h*2}) in terms of conformal time $\eta\equiv\int_0^t\frac{dt'}{R(t')}$, and defining $\bar{h}_{i}=Rh_i$ $(i=+,\times)$, the equations (\ref{eq h+2},\ref{eq h*2}) are simplified as
\begin{eqnarray}\label{bar h+}
\frac{d^2\bar{h}_+}{d\eta^2}+\left(k^2_3-\frac{1}{R}\frac{d^2R}{d\eta^2}\right)\bar{h}_+&=&-\frac{3bR}{2}k_1^2,\\
\label{bar h*}
\frac{d^2\bar{h}_\times}{d\eta^2}+\left(k^2_3-\frac{1}{R}\frac{d^2R}{d\eta^2}\right)\bar{h}_\times&=&0.
\end{eqnarray}
During the inflation period of the universe, the scale factor $R(t)$ expands nearly exponentially, i.e.,$R(t)\sim \exp(Ht)$. Then, the relations $\eta\simeq-\frac{1}{RH}$ and $\frac{1}{R}\frac{d^2R}{d\eta^2}\simeq\frac{2}{\eta^2}$ hold approximately. By making use of these approximate relations, the equations (\ref{bar h+}) and (\ref{bar h*}) become
\begin{eqnarray}\label{bar h+1}
\frac{d^2\bar{h}_+}{d\eta^2}+\left(k^2_3-\frac{2}{\eta^2}\right)\bar{h}_+&=&\frac{3b}{2H}\frac{k_1^2}{\eta},\\
\label{bar h*1}
\frac{d^2\bar{h}_\times}{d\eta^2}+\left(k^2_3-\frac{2}{\eta^2}\right)\bar{h}_\times&=&0.
\end{eqnarray}
Noticing that $b$ depends only on $\vec{x}$ and $H$ is nearly equal to a constant during inflation, we find the solution of the equations (\ref{bar h+1}) and (\ref{bar h*1}) that
\begin{eqnarray}
\bar{h}_+&=&\frac{e^{-ik_3\eta}}{\sqrt{2k_3}}\left(1-\frac{i}{k_3\eta}\right)+\frac{3b}{2H}\frac{k_1^2}{k^2_3},\\
\bar{h}_\times&=&\frac{e^{-ik_3\eta}}{\sqrt{2k_3}}\left(1-\frac{i}{k_3\eta}\right).
\end{eqnarray}

The primordial power spectrum is defined by
\begin{eqnarray}
\mathcal{P}_h\equiv\frac{k^3}{2\pi^2}|h(\vec{k})|^2.
\end{eqnarray}
Then, the primordial power spectrum of $h_+$ and $h_\times$ are given as
\begin{eqnarray}\label{P h+}
\mathcal{P}_{h_+}&=&\frac{H^2}{(2\pi)^2}(1+k^2_3\eta^2)+\frac{3bH}{2\sqrt{2}\pi^2}\sqrt{k_3}(k_1\eta)^2\left(\cos(k_3\eta)-\frac{\sin(k_3\eta)}{k_3\eta}\right),\\
\mathcal{P}_{h_\times}&=&\frac{H^2}{(2\pi)^2}(1+k^2_3\eta^2),
\end{eqnarray}
where we have omitted the second order terms in $\mathcal{P}_{h_+}$. For the super-horizon perturbations ($k_3\eta\ll1$), both $\mathcal{P}_{h_+}$ and $\mathcal{P}_{h_\times}$ equal to $\frac{H^2}{(2\pi)^2}$, which are the same with those in the $\Lambda$CDM model. For the sub-horizon perturbations ($k_3\eta\gg1$), $\mathcal{P}_{h_+}$ reduces to
\begin{eqnarray}\label{P h+2}
\mathcal{P}_{h_+}=\frac{H^2}{(2\pi)^2}k_3^2\eta^2+\frac{3bH}{2\sqrt{2}\pi^2}\sqrt{k_3}(k_1\eta)^2\cos(k_3\eta).
\end{eqnarray}

The above discussions are given under the condition that the Finslerian perturbation $b$ is a function of spatial coordinate $\vec{x}$ only. If $Ah_+(t)=b(t)$ in Fourier space, following the same process, we obtain the power spectrum of super-horizon perturbation $h_+$
\begin{eqnarray}
\mathcal{P}_{h_+}&=&\frac{H^2}{(2\pi)^2}\left(\frac{k_3}{\sqrt{k^2_3+\frac{3A}{2}k_1^2}}\right)^3\nonumber\\
\label{P h+1}
&\simeq&\frac{H^2}{(2\pi)^2}\left(1+\frac{9A}{4}\left(\frac{k_1}{k_3}\right)^2\right),
\end{eqnarray}
where $A$ denotes the magnitude of $b$.


Also, the Finslerian perturbation $b$ makes the gravitational field equation (\ref{eq h+1}) differ from (\ref{eq h*1}). So, it could change the polarization of the gravitational wave.

\section{Conclusions and Remarks}
In this paper, we have presented a Finslerian perturbation for the $\Lambda$CDM model. Instead of studying the gravitational behaviors in Riemann spacetime, our discussions are based on the Randers-Finsler spacetime (\ref{Randers form}). The data set of WMAP and Planck Collaboration hint that our universe may have a preferred direction. As a generalization of Riemann spacetime, Finsler spacetime
involves a preferred direction naturally. We have shown that a Finslerian perturbation (\ref{one form}) breaks the isotropic symmetry of the FRW metric.

During the thermal history of the universe, the standard model for elementary particles functions well. So, the Finsler spacetime should not modify the dispersion relations of particles during this era. It implies that no Finslerian effect exists in the collisions of particles. But, the Finslerian perturbation changes the geodesic. It will affect the term $\frac{df}{dt}$ in Boltzmann equations. Our investigation showed that the Boltzmann equation for massless particles in Randers-Finsler spacetime is the same with the one \cite{Dodelson} in the FRW metric with scalar perturbations $\psi$ and $\phi$. As for massive particles, one can simply get Boltzmann equations in Randers-Finsler spacetime by transforming $\psi\longrightarrow\psi+b$ from the one in FRW metric with scalar perturbations.

Up to now, debate still exists on the gravitational field equation in Finsler spacetime. There are many types of gravitational field equation in Finsler spacetime \cite{Miron,Rutz,Vacaru,Pfeifer}. And these equations do not equal to each other. The Finslerian length element $F$ is constructed on a tangent bundle, so does the curvature tensor and the corresponded energy-momentum tensor in the gravitational field equation. However, the energy-momentum tensor which is constructed on a tangent bundle is rather obscure in physics. In this paper, we have proposed a perturbational version of gravitational field equation (\ref{Gravitation eq}) in Finsler spacetime, where we have omitted the curvature tensor that does not belong to the base space of the tangent bundle. We have shown that the equation (\ref{Gravitation eq}) could be derived from the formulism of Pfeifer {\it et al}. \cite{Pfeifer} approximately. And the equation (\ref{Gravitation eq}) was constructed by the geometrical invariant, namely, the Ricci tensor, which means that it is insensitive to the connections. Thus, our gravitational field equation is universal and meet the physical requirement. In Randers-Finsler spacetime(\ref{CMB a00},\ref{CMB aij},\ref{one form}), we have presented the gravitational field equations (\ref{Gravitation eq G00},\ref{Gravitation eq G0j},\ref{Gravitation eq Gij}) with the Finslerian perturbation. Certainly, the Finslerian perturbation parameter $b$ need be constrained by the data of WMAP and Planck satellite.

The gravitational field equations (\ref{eq h+1},\ref{eq h*1}) of the gravitational wave were given. To involve the Finslerian effect in the gravitational wave, we have shown that the wave vector of the Finslerian perturbation parameter $b$ should perpendicular to the wave vector of the gravitational wave. In Rander-Finsler spacetime, the primordial power spectrum of the gravitational wave was investigated. We have shown that the primordial power spectrum for super-horizon perturbations is unchanged. But, for sub-horizon perturbations, the power spectrum is changed as (\ref{P h+2}).

\vspace{1cm}
\begin{acknowledgments}
We would like to thank  Y. G. Jiang and H.N. Lin for useful discussions. Project 11375203 and 11305181 supported by NSFC.
\end{acknowledgments}

\end{document}